\documentclass[aps,pra,showpacs,superscriptaddress,preprint]{revtex4}
\usepackage{graphicx}

\catcode`ð=\active
 \defð{\u{g}}
 \catcode`Ð=\active
 \defÐ{\u{G}}
 \catcode`Ý=\active
 \defÝ{\. I}
 \catcode`ö=\active
 \defö{\"{o}}
 \catcode`Ö=\active
 \defÖ{\"O}
 \catcode`ü=\active
 \defü{\"{u}}
 \catcode`Ü=\active
 \defÜ{\"{U}}
 \catcode`Þ=\active
 \defÞ{\c{S}}
 \catcode`þ=\active
 \defþ{\c{s}}
 \catcode`ý=\active
 \defý{{\i}}
 \catcode`ç=\active
 \defç{\d{c}}
 \catcode`Ç=\active
 \defÇ{\d{C}}

\begin{document}

\title{Approximate Analytical Solutions of the Klein-Gordon Equation for Hulth\'{e}n
Potential with Position-Dependent Mass}

\author{\small Altuð Arda}
\email[E-mail: ]{arda@hacettepe.edu.tr}\affiliation{Department of
Physics Education, Hacettepe University, 06800, Ankara,Turkey}
\author{\small Ramazan Sever}
\email[E-mail: ]{sever@metu.edu.tr}\affiliation{Department of
Physics, Middle East Technical  University, 06800, Ankara,Turkey}
\author{\small Cevdet Tezcan}
\email[E-mail: ]{ctezcan@baskent.edu.tr}\affiliation{Faculty of
Engineering, Baþkent University, Baglýca Campus, Ankara,Turkey}

\date{\today}

\begin{abstract}
The Klein-Gordon equation is solved approximately for the Hulth\'{e}n
potential for any angular momentum quantum number $\ell$ with the
position-dependent mass. Solutions are obtained reducing the
Klein-Gordon equation into a Schr\"{o}dinger-like differential
equation by using an appropriate coordinate transformation. The
Nikiforov-Uvarov method is used in the calculations to get an energy
eigenvalue and and the wave functions. It is found that the results
in the case of constant
mass are in good agreement with the ones obtained in the literature.\\
Keywords: Hulth\'{e}n potential, Klein-Gordon equation,
Position-Dependent Mass, Nikiforov-Uvarov Method
\end{abstract}
\pacs{03.65.-w; 03.65.Ge; 12.39.Fd}

\maketitle

\newpage

\section{Introduction}
Exact or approximate solutions of the relativistic/non-relativistic
wave equations have received great attentions. So far the solutions
are in general obtained for the case of constant mass or at most
time-dependent mass [1, 2]. The effective mass solutions have
received much attentions recently. A quite general hermitian
effective Hamiltonian is used to describe the non-relativistic
systems, such description is applied to study the semiconductor
nanostructures [3]. Another interesting problem is that the correct
form of the kinetic energy operator for such a Hamiltonian, since
the momentum, and the mass operators are no longer commute in the
case of position-dependent mass, which is related to the problem of
ordering ambiguity [4]. There are some important problems related to
the ordering ambiguity concept, such as the dependence of nuclear
forces on the relative velocity of the two nucleons [5, 6], the
impurities of crystals [7]. In addition, many authors have studied
to propose some effective Hamiltonians for non-relativistic case
taking into account the dependence of the mass on position [8].

There are many efforts about solving the Schr\"{o}dinger equation for
the case of position-dependent mass by using different methods or
schemes for different potentials, such as exponential type potential
[4], Natanzon potentials by using a group-theoretical method [9],
solutions in the case of mappings of the Morse+oscillator+Coulomb
potential [10], hyperbolic-type potentials [11], Morse, and Coulomb
potential with the position-dependent mass [12, 13], $PT$-symmetric
anharmonic oscillators [14], the Morse-like potential in the scheme
of supersymmetric quantum mechanics [15], Kratzer and Scarf II
potentials [16], deformed Rosen-Morse, and Scarf potentials [17].
Many authors have been also solved the Klein-Gordon, and Dirac
equation by taking a suitable mass distributions in one and/or three
dimensional cases for different potentials, such as Coulomb
potential [18], Lorentz scalar interactions[19], hyperbolic-type
potentials [20], Morse potential [21], and P\"{o}schl-teller potential
[22].

Here we intend to solve the Klein-Gordon equation within the
framework of an approximation to the centrifugal potential term. We
study the effect of the mass varying with position on the energy
spectra, and the eigenfunctions of the vector, and scalar Hulth\'{e}n
potential [23], which is widely used in nuclear, particle physics,
atomic physics, condensed matter, and chemical physics [24-26]. For
our task, we use a general parametric form of the Nikiforov-Uvarov
(NU) method, which is based on turning of a second order
differential equations to a hypergeometric type equation [27].

The organization of this work is as follows. In Section II, we
give briefly the parametric generalization of the NU-method. In
Section III, we give the energy eigenvalue equation, and
corresponding eigenfunctions for the vector, scalar Hulth\'{e}n
potential for any $\ell$-values in the position-dependent mass
background. We obtain also the results for the case of the
constant mass, and we summarize our concluding in Section IV.

\section{Nikiforov-Uvarov Method}

The Schr\"{o}dinger equation can be transformed into a second
order differential equation with the following form

\begin{eqnarray}
\sigma^{2}(s)\frac{d^{2}\Psi(s)}{ds^{2}}+\sigma(s)\tilde{\tau}(s)
\frac{d\Psi(s)}{ds}+\tilde{\sigma}(s)\Psi(s)=0\,,
\end{eqnarray}
where $\sigma(s)$, $\tilde{\sigma}(s)$ are polynomials, at most,
second degree, and $\tilde{\tau}(s)$ is a first degree polynomial.
In order to find a particular solution, we take the following form

\begin{eqnarray}
\Psi(s)=\psi(s)~\phi(s),
\end{eqnarray}
We get from Eq. (1)

\begin{eqnarray}
\sigma(s)\frac{d^2\phi(s)}{ds^2}+\tau(s)\frac{d\phi(s)}{ds}+\lambda\phi(s)=0\,,
\end{eqnarray}
where $\phi(s)$ can be written in terms of Rodriguez formula

\begin{eqnarray}
\phi_{n}(s)=\frac{B_{n}}{\rho(s)}\frac{d^{n}}{ds^{n}}
\left[\sigma^{n}(s)~\rho(s)\right],
\end{eqnarray}
and the weight function $\rho(s)$ satisfies

\begin{eqnarray}
\frac{d\sigma(s)}{ds}+\frac{\sigma(s)}{\rho(s)}\frac{d\rho(s)}{ds}=\tau(s)\,.
\end{eqnarray}

The other factor of the solution is defined as

\begin{eqnarray}
\frac{1}{\psi(s)}\frac{\psi(s)}{ds}=\frac{\pi(s)}{\sigma(s)}\,.
\end{eqnarray}
In the method, the polynomial $\pi(s)$, and the parameter $k$ are
defined as [27]

\begin{eqnarray}
\pi(s)=\frac{1}{2}\,[\sigma'(s)-\tilde{\tau}(s)]\pm\Big\{\frac{1}{4}
[\sigma'(s)-\tilde{\tau}(s)]^2-\tilde{\sigma}(s)+k\sigma(s)\Big\}^{1/2},
\end{eqnarray}
and

\begin{eqnarray}
\lambda=k+\pi^{\prime}(s )\,.
\end{eqnarray}
where $\lambda$ is a constant, and given in Eq. (3). Since square
root in the polynomial $\pi(s)$ in Eq. (7) must be a square then
this defines the constant $k.$ Replacing $k$ into Eq. (7), we
define

\begin{eqnarray}
\tau(s)=\tilde{\tau}(s)+2\pi(s).
\end{eqnarray}
Since $\rho(s) > 0$\, and $\sigma(s) > 0$, hence the derivative of
$\tau(s)$ should be negative [27], which leads to the choice of
the solution. If $\lambda$ in Eq. (8) is

\begin{eqnarray}
\lambda=\lambda_{n}=-n\tau^{\prime}-\frac{\left[n(n-1)\sigma^{\prime\prime}\right]}{2},
\quad n=0,1,2,\ldots
\end{eqnarray}
the hypergeometric type equation has a particular solution with
degree $n$.

In order to explain the general parametric form of the NU method,
let us take the general form of a Schr\"{o}dinger-like equation
including any potential

\begin{eqnarray}
[s(1-\alpha_{3}s)]^{2}\frac{d^{2}\Psi(s)}{ds^{2}}+[s(1-\alpha_{3}s)(\alpha_{1}-\alpha_{2}s)]
\frac{d\Psi(s)}{ds}+[-\xi_{1}s^{2}+\xi_{2}s-\xi_{3}]\Psi(s)=0.
\end{eqnarray}
When Eq. (11) is compared with Eq. (1), we get

\begin{eqnarray}
\tilde{\tau}(s)=\alpha_{1}-\alpha_{2}s\,\,;\,\sigma(s)=s(1-\alpha_{3}s)\,\,;\,
\tilde{\sigma}(s)=-\xi_{1}s^{2}+\xi_{2}s-\xi_{3}.
\end{eqnarray}

Substituting these into Eq. (7), we get

\begin{eqnarray}
\pi(s)=\alpha_{4}+\alpha_{5}s\pm\sqrt{(\alpha_{6}-k\alpha_{3})s^{2}+(\alpha_{7}+k)s+\alpha_{8}}\,,
\end{eqnarray}
where the parameters in the above equation are as follows

\begin{eqnarray}
\begin{array}{ll}
\alpha_{4}=\frac{1}{2}\,(1-\alpha_{1})\,, & \alpha_{5}=\frac{1}{2}\,(\alpha_{2}-2\alpha_{3})\,, \\
\alpha_{6}=\alpha_{5}^{2}+\xi_{1},\,, &
\alpha_{7}=2\alpha_{4}\alpha_{5}-\xi_{2}\,, \\
\alpha_{8}=\alpha_{4}^{2}+\xi_{3}\,. & \\
\end{array}
\end{eqnarray}

In NU-method, the function under square root must be the square of
a polynomial, so

\begin{eqnarray}
k_{1,2}=-(\alpha_{7}+2\alpha_{3}\alpha_{8})\pm2\sqrt{\alpha_{8}\alpha_{9}},
\end{eqnarray}
where

\begin{eqnarray}
\alpha_{9}=\alpha_{3}\alpha_{7}+\alpha_{3}^{2}\alpha_{8}+\alpha_{6}.
\end{eqnarray}
The function $\pi(s)$ becomes

\begin{eqnarray}
\pi(s)=\alpha_{4}+\alpha_{5}s-\left[(\sqrt{\alpha_{9}}+\alpha_{3}
\sqrt{\alpha_{8}})s-\sqrt{\alpha_{8}}\,\right].
\end{eqnarray}
for the $k$-value
$k=-(\alpha_{7}+2\alpha_{3}\alpha_{8})-2\sqrt{\alpha_{8}\alpha_{9}}$\,,
where we have to say that the different $k$'s lead to the
different $\pi(s)$'s. We also have from Eq. (9)

\begin{eqnarray}
\tau(s)=\alpha_{1}+2\alpha_{4}-(\alpha_{2}-2\alpha_{5})s-2\left[(\sqrt{\alpha_{9}}
+\alpha_{3}\sqrt{\alpha_{8}}\,)s-\sqrt{\alpha_{8}}\,\right].
\end{eqnarray}
Thus, we impose the following for satisfying the condition that
the derivative of $\tau(s)$ must be negative

\begin{eqnarray}
\tau^{\prime}(s)&=&-(\alpha_{2}-2\alpha_{5})-2(\sqrt{\alpha_{9}}
+\alpha_{3}\sqrt{\alpha_{8}}\,) \nonumber \\[5mm]
                              &=&-2\alpha_{3}-2(\sqrt{\alpha_{9}}+\alpha_{3}\sqrt{\alpha_{8}}\,)\quad
                              <0.
\end{eqnarray}
From Eqs. (8), (9), (18), and (19), and equating Eq. (8) with the
condition that $\lambda$ should satisfy given by Eq. (10), we
obtain

\begin{eqnarray} \label{eq.29}
\alpha_{2}n-(2n+1)\alpha_{5}&+&(2n+1)(\sqrt{\alpha_{9}}+\alpha_{3}
\sqrt{\alpha_{8}}\,)+n(n-1)\alpha_{3}\nonumber \\[5mm]
     &+&\alpha_{7}+2\alpha_{3}\alpha_{8}+2\sqrt{\alpha_{8}\alpha_{9}}=0.
\end{eqnarray}
which is the energy eigenvalue equation of a given potential.

Now, let us look the eigenfunctions of the problem with any
potential. We obtain the second part of the solution from Eq. (4)

\begin{eqnarray}
\phi_{n}(s)=P_{n}^{(\alpha_{10}-1,\frac{\alpha_{11}}{\alpha_{3}}-\alpha_{10}-1)}(1-2\alpha_{3}s)\,,
\end{eqnarray}
by using the explicit form of the weight function obtained from
Eq. (5)

\begin{eqnarray}
\rho(s)=s^{\alpha_{10}-1}(1-\alpha_{3}s)^{\frac{\alpha_{11}}{\alpha_{3}}-\alpha_{10}-1}\,,
\end{eqnarray}
where

\begin{eqnarray}
\alpha_{10}=\alpha_{1}+2\alpha_{4}+2\sqrt{\alpha_{8}}\,\,\,;\,\,
\alpha_{11}=\alpha_{2}-2\alpha_{5}+2(\sqrt{\alpha_{9}}+\alpha_{3}\sqrt{\alpha_{8}}\,)\,.
\end{eqnarray}
and $P_{n}^{(\alpha,\beta)}(1-2\alpha_{3}s)$ are Jacobi
polynomials. From Eq. (6), one gets

\begin{eqnarray}
\psi(s)=s^{\alpha_{12}}(1-\alpha_{3}s)^{-\,\alpha_{12}-\frac{\alpha_{13}}{\alpha_{3}}}\,,
\end{eqnarray}
then the general solution $\Psi(s)=\psi(s)\phi(s)$ becomes

\begin{eqnarray}
\psi(s)=s^{\alpha_{12}}(1-\alpha_{3}s)^{-\alpha_{12}-\frac{\alpha_{13}}{\alpha_{3}}}
P_{n}^{(\alpha_{10}-1,\frac{\alpha_{11}}{\alpha_{3}}-\alpha_{10}-1)}(1-2\alpha_{3}s)\,,
\end{eqnarray}
where

\begin{eqnarray}
\alpha_{12}=\alpha_{4}+\sqrt{\alpha_{8}}\,\,\,;\,\,
\alpha_{13}=\alpha_{5}-(\sqrt{\alpha_{9}}+\alpha_{3}\sqrt{\alpha_{8}}\,).
\end{eqnarray}

\section{Bound-State Solutions}
The Klein-Gordon equation for a particle with mass $m$ with vector
$V_v(r)$, and scalar $V_s(r)$ potentials is ($\hbar=c=1$)

\begin{eqnarray}
\Big\{-\nabla^2-[E^2-m^2(r)]+2[m(r)V_s(r)+EV_v(r)]+[V^2_s(r)-V^2_v(r)]\Big\}\Psi(r,\theta,\phi)=0\,,
\end{eqnarray}
Using $\Psi(r,\theta,\phi)=r^{-1}\phi(r)Y_{\ell\,m}(\theta,\phi)$,
we have the radial part o the equation

\begin{eqnarray}
\frac{d^2\phi(r)}{dr^2}&+&\Big\{[E^2-m^2(r)]-2[m(r)V_s(r)+EV_v(r)]\nonumber\\&+&[V^2_s(r)-V^2_v(r)]
-\,\frac{\ell(\ell+1)}{r^2}\,\Big\}\phi(r)=0\,.
\end{eqnarray}
where $Y_{\ell\,m}(\theta,\phi)$ is spherical harmonics, and $\ell$ is the angular momentum
quantum number.

In order to solve the Eq. (30), we prefer to use the following
mass function

\begin{eqnarray}
m(r)=m_0+\,\frac{m_1e^{-r/r_0}}{1-e^{-r/r_0}}\,,
\end{eqnarray}
where $m_0$, $m_1$ are two arbitrary, positive constants. We have
to use an approximation, given by
$1/r^2\approx\,e^{r/r_0}/(e^{r/r_0}-1)^2r^2_0$, to the centrifugal term, since the radial
equation has no analytical solutions for $\ell\neq0$ [28, 29]. By
taking the scalar, and vector potentials as the Hulth\'{e}n potential

\begin{eqnarray}
V_s(r)=-\,\frac{S_0}{e^{r/r_0}-1}\,\,;\,\,\,V_v(r)=-\,\frac{V_0}{e^{r/r_0}-1}\,,
\end{eqnarray}
and using the Eq. (31), we get

\begin{eqnarray}
\frac{d^2\phi(r)}{dr^2}+\Big\{E^2-m^2_0&+&\,\frac{2m_0(S_0-m_1)+2EV_0}{e^{r/r_0}-1}\,+\,
\frac{2m_1
S_0-m^2_1+V^2_0-S^2_0}{(e^{r/r_0}-1)^2}\nonumber\\
&-&\frac{\ell(\ell+1)e^{r/r_0}}{r^2_0(e^{r/r_0}-1)^2}\Big\}\phi(r)=0
\end{eqnarray}

By using a new variable $e^{-r/r_0}=s$, Eq. (33) becomes

\begin{eqnarray}
\frac{d^2\phi(s)}{ds^2}&+&\,\frac{1-s}{s(1-s)}\,\frac{d\phi(s)}{ds}\,+
\Bigg\{\frac{r^2_0(E^2-m^2_0)}{s^2}\,+\,\frac{2r^2_0(m_0(S_0-m_1)+EV_0)}{s(1-s)}\nonumber\\
&+&\frac{r^2_0(m_1(2S_0-m_1)+V^2_0-S^2_0)}{(1-s)^2}\,-\,\frac{\ell(\ell+1)}{s(1-s)^2}\Bigg\}\phi(s)=0\,,
\end{eqnarray}

By using the new parameters

\begin{eqnarray}
\alpha(m_1)&=&\eta(m_1)r_0\,,\nonumber\\
\eta^2(m_1)&=&(m_1-m_0)^2-E^2\,,\nonumber\\
\beta^2_{1}(m_1)&=&r^2_0[2EV_0-2S_0(m_1-m_0)]\,,\nonumber\\
\beta^2_{2}(m_1)&=&r^2_0[2EV_0-2m_0(m_1-S_0)]\,,\nonumber\\
\nu^2(m_1)&=&-\alpha^2+\alpha^2(m_1)+\beta^2_{1}(m_1)-\beta^2_{2}(m_1)+\nu^2\,,
\end{eqnarray}
where $\nu(m_1)(m_1 \rightarrow 0)=\nu$\,,\,$\eta(m_1)(m_1
\rightarrow 0)=\eta$, and $\alpha=\eta r_0$, and comparing Eq.
(34) with Eq. (11), we get the following parameter set given in
Section II

\begin{eqnarray}
\begin{array}{ll}
\alpha_1=1\,, & \xi_1=\alpha^2(m_1)+\beta^2(m_1)+\nu^2(m_1) \\
\alpha_2=1\,, &
\xi_2=2\alpha^2+\beta^2_{2}(m_1)-\ell(\ell+1) \\
\alpha_3=1\,, &
\xi_3=\alpha^2 \\ \alpha_4=0\,, & \alpha_5=-\,\frac{1}{2} \\
\alpha_6=\xi_1+\frac{1}{4}\,, & \alpha_7=-\xi_2 \\
\alpha_8=\xi_3\,, & \alpha_9=\xi_1-\xi_2+\xi_3+\frac{1}{4} \\
\alpha_{10}=1+2\sqrt{\xi_3}\,, & \alpha_{11}=2+2(\,\sqrt{\xi_1-\xi_2+\xi_3+\frac{1}{4}\,}+\sqrt{\xi_3}\,) \\
\alpha_{12}=\sqrt{\xi_3}\,, &
\alpha_{13}=-\frac{1}{2}-(\,\sqrt{\xi_1-\xi_2+\xi_3+\frac{1}{4}\,}+\sqrt{\xi_3}\,)
\end{array}
\end{eqnarray}
where $\nu^2=r^2_0(S^2_0-V^2_0)$\,, and $\eta^2=m^2_0-E^2$ in the above equations.

We can easily get the energy eigenvalue equation of the Hulth\'{e}n
potential by using Eq. (20)

\begin{eqnarray}
\alpha=\,\frac{\beta^2_{2}(m_1)-\ell(\ell+1)-n^2-(2n-1)\delta\,'}{2(n+\delta\,')}\,,
\end{eqnarray}
where
$\delta\,'=\frac{1}{2}+\frac{1}{2}\sqrt{(2\ell+1)^2+4\nu^2(m_1)\,}$\,.
We list some energy eigenvalues in Table I, and Table II for the
case of constant mass, and the one of spatially dependent mass,
respectively. To compare our results, we have used the values of the
parameters given in Ref. [32]. $E_{a}$ denotes the energy
eigenvalues of the particle, and $E_{p}$ denotes the one of the
antiparticle in Table I, and Table II.

According the result obtained in Eq. (37), we give easily the eigenvalue equation in
the case of constant mass

\begin{eqnarray}
\alpha=\,\frac{\beta^2_{2}(m_1=0)-\ell(\ell+1)-n^2-(2n-1)\delta\,'(m_1=0)}{2(n+\delta\,'(m_1=0))}\,,
\end{eqnarray}
which is the same with the result obtained in Ref. [28].

The corresponding eigenfunctions of the Hulth\'{e}n potential is
written by using Eq. (26), and Eq. (35)

\begin{eqnarray}
\phi(r)=A_{n}e^{-\alpha r/r_0}\,(1-e^{-r/r_0})^{1+\delta\,'}\,P_{n}^{(2\alpha\,,\,1+2\delta\,')}(1-2e^{-r/r_0})\,,
\end{eqnarray}
where $A_{n}$ is a normalization constant.

Finally, the eigenfunctions in the case of constant mass are
written by using Eq. (39)

\begin{eqnarray}
\phi(r)=A'_{n}e^{-\alpha r/r_0}\,(1-e^{-r/r_0})^{1+\delta\,''}\,P_{n}^{(2\alpha\,,\,1+2\delta\,'')}(1-2e^{-r/r_0})\,.
\end{eqnarray}
where
$\delta\,''=\frac{1}{2}+\frac{1}{2}\sqrt{(2\ell+1)^2+4\nu^2(m_1=0)\,}$\,.
The Jacobi polynomials
$P_{n}^{(2\alpha\,,\,1+2\delta\,'')}(1-2e^{-r/r_0})$ in the last
result can be written in terms of hypergeometric function
$_2F_1(-n,n+2\alpha+2\delta\,'+2,2\alpha;s)$\,, which gives the
same result obtained in Ref. [28].

The normalization constant in Eq. (39) is obtained from the
the normalization condition

\begin{eqnarray}
\int_{0}^{\infty}|\phi(r)|\,^2dr=1\,,
\end{eqnarray}

By introducing a new variable as $x=1-2e^{-r/r_{0}}$, we have from Eq. (41)

\begin{eqnarray}
|A_{n}|^2\frac{r_{0}}{2^{1+2\alpha+\beta}}\int_{-1}^{+1}(1-x)^{2\alpha-1}(1+x)(1+x)^{\beta}
P_{n}^{(2\alpha\,,\,\beta)}(x)P_{m}^{(2\alpha\,,\,\beta)}(x)=1\,,
\end{eqnarray}
where $\beta=1+2\delta\,'$\,. By using the following identities [30, 31]

\begin{eqnarray}
2n(\zeta+\zeta'+n)(\zeta+\zeta'+2n-2)P_{n}^{(\zeta\,,\,\zeta')}(x)
=(\zeta+\zeta'+2n-1)(\zeta^2-\zeta'^2)P_{n-1}^{(\zeta\,,\,\zeta)}(x)\nonumber\\+(\zeta+\zeta'+2n-1)(\zeta+\zeta'+2n)
(\zeta+\zeta'+2n-2)xP_{n-1}^{(\zeta\,,\,\zeta')}(x)\nonumber\\-2(\zeta+n-1)(\zeta'+n-1)(\zeta+\zeta'+2n)
P_{n-2}^{(\zeta\,,\,\zeta')}(x)\,,
\end{eqnarray}
and

\begin{eqnarray}
\int_{-1}^{+1}(1-x)^{\zeta-1}(1+x)^{\zeta'}
[P_{n}^{(\zeta\,,\,\zeta')}(x)]^2dx=\frac{2^{\zeta+\zeta'}\Gamma(\zeta+n+1)\Gamma(\zeta'+n+1)}
{n!\zeta\Gamma(\zeta+\zeta'+n+1)}\,,
\end{eqnarray}
we obtain the normalization constant

\begin{eqnarray}
A_{n}=\frac{2}{\sqrt{r_{0}\,}}\sqrt{n!\alpha\frac{(2\alpha+\beta+2n+2)(2\alpha+\beta+2n)}{4n(n+1+2\alpha+\beta)
+2(1+\beta)(2\alpha+\beta)}\frac{\Gamma(2\alpha+\beta+n+1)}{\Gamma(2\alpha+n+1)\Gamma(\beta+n+1)}\,}\,.
\end{eqnarray}

By following the same procedure, the normalization constant
$A'_{n}$ in the eigenfunctions of the case of constant mass is
obtained as $A'_{n}=A_{n}(\beta \rightarrow 1+2\delta\,'')$ in Eq.
(43).

\section{Conclusion}
We have approximately solved the Klein-Gordon equation for the
Hulth\'{e}n potential for any angular momentum quantum number in the
position-dependent mass background. We have found the eigenvalue
equation, and corresponding wave functions in terms of Jacobi
polynomials by using NU-method within the framework of an
approximation to the centrifugal potential term. We have also
obtained the energy eigenvalue equation, and corresponding
eigenfunctions for the case of the constant mass. Results for the
case of constant mass are the same with the ones obtained in Ref.
[28].

\section{Acknowledgments}
This research was partially supported by the Scientific and
Technical Research Council of Turkey.

\newpage

\begin{table}
\caption{The energy eigenvalues of vector, and scalar H\'{u}lthen potential
for $m_{0}=1$\,, and $m_{1}=0$.}
\begin{ruledtabular}
\begin{tabular}{cccccccc}
$V_{0}=S_{0}=1$ &  &  & & &  & & \\
\hline
$n$ & $\ell$ & $E_{a}\footnotemark[1]$ & $E_{p}\footnotemark[1]$ & $E_{a}\footnotemark[2]$ & $E_{p}\footnotemark[2]$ &
$E_{a}$\footnotemark[3] & $E_{p}\footnotemark[3]$\\
1 & 0 & -0.6000000 & 1.0000000 & -0.6000000 & 1.0000000 & -0.6000000 & 1.0000000\\
1 & 1 & --- & --- & --- & --- & --- & ---\\
$V_{0}=S_{0}=2$ &  &  & & &  & & \\ \hline
1 & 0 & -0.7071068 & 0.7071068 & -0.7071068 & 0.7071068 & -0.7071068 & 0.7071068\\
1 & 1 & -0.2149407 & 0.9841714 & --- & --- & --- & ---\\
2 & 0 & -0.2149407 & 0.9841714 & -0.2149410 & 0.9841710 & -0.2149410 & 0.9841710\\
2 & 1 & --- & --- & --- & --- & --- & ---\\
2 & 2 & --- & --- & --- & --- & --- & ---\\
$V_{0}=S_{0}=3$ &  &  & & &  & & \\ \hline
1 & 0 & -0.7637079 & 0.3021695 & -0.7637080 & 0.3021690 & -0.7637080 & 0.3021690\\
1 & 1 & -0.4114378 & 0.9114378 & --- & --- & --- & ---\\
2 & 0 & -0.4114378 & 0.9114378 & -0.4114380 & 0.9114380 & -0.4114380 & 0.9114380\\
2 & 1 & 0.6000000 & 0.6000000 & --- & --- & --- & ---\\
2 & 2 & --- & --- & --- & --- & --- & ---\\
3 & 0 & 0.6000000 & 0.6000000 & 0.6000000 & 0.6000000 & 0.6000000 & 1.0000000\\
$V_{0}=S_{0}=6$ &  &  & & &  & & \\ \hline
1 & 0 & -0.8449490 & -0.3550510 & -0.8449490 & -0.3550510 & -0.8449490 & -0.3550510\\
1 & 1 & -0.6358899 & 0.2358899 & --- & --- & --- & ---\\
2 & 0 & -0.6358899 & 0.2358899 & -0.6358900 & 0.2358900 & -0.6358900 & 0.2358900\\
2 & 1 & -0.3021695 & 0.7637079 & --- & --- & --- & ---\\
2 & 2 & 0.2844158 & 0.9942727 & --- & --- & --- & ---\\
3 & 0 & -0.3021695 & 0.7637079 & -0.3021690 & 0.7637080 & -0.3021690 & 0.7637080\\
3 & 1 & 0.2844158 & 0.9942727 & --- & --- & --- & ---\\
4 & 0 & 0.2844158 & 0.9942727 & 0.284416 & 0.994273 & 0.2844160 & 0.9942730\\
\end{tabular}
\end{ruledtabular}
\footnotetext[1]{our results} \footnotetext[2]{results obtained in
Ref. [32]} \footnotetext[3]{results obtained in
Ref. [33], and Ref. [34]}
\end{table}

\newpage

\begin{table}
\caption{The energy eigenvalues of vector, and scalar H\'{u}lthen potential
for $m_{1}\neq 0$.}
\begin{ruledtabular}
\begin{tabular}{cccccccc}
$m_{1}$ & $m_{0}$ & $V_{0}$ & $S_{0}$ & $n$ & $\ell$ &
$E_{a}$ & $E_{p}$\\ \hline
0.1 & 5 & 1 & 1 & 1 & 0 & -4.868720 & 3.443410\\
 &  &  &  & 1 & 1 & -4.742880 & 4.722690\\
 &  &  &  & 2 & 0 & -4.768190 & 4.618770\\
 &  &  &  & 2 & 1 & -4.577550 & 4.982510\\
 &  &  &  & 2 & 2 & -4.347700 & 4.964780\\
 &  &  &  & 3 & 0 & -4.613290 & 4.960360\\
 &  &  &  & 3 & 1 & -4.354450 & 4.967570\\
 &  &  &  & 3 & 2 & -4.056980 & 4.788530\\
 &  &  &  & 3 & 3 & -3.682040 & 4.484330\\ \hline
0.01 & 5 & 2 & 2 & 1 & 0 & -4.913410 & 0.8229250\\
 &  &  &  & 1 & 1 & -4.804170 & 3.110670\\
 &  &  &  & 2 & 0 & -4.807820 & 3.065630\\
 &  &  &  & 2 & 1 & -4.650830 & 4.252020\\
 &  &  &  & 2 & 2 & -4.445800 & 4.795730\\
 &  &  &  & 3 & 0 & -4.655840 & 4.229630\\
 &  &  &  & 3 & 1 & -4.447040 & 4.793910\\
 &  &  &  & 3 & 2 & -4.185200 & 4.989330\\
 &  &  &  & 3 & 3 & -3.857960 & 4.956220\\ \hline
0.1 & 5 & -1 & 1 & 1 & 0 & -3.443410 & 4.868720\\
 &  &  &  & 1 & 1 & -4.722690 & 4.742880\\
 &  &  &  & 2 & 0 & -4.618770 & 4.768190\\
 &  &  &  & 2 & 1 & -4.982510 & 4.577550\\
 &  &  &  & 2 & 2 & -4.964780 & 4.347700\\
 &  &  &  & 3 & 0 & -4.960360 & 4.613920\\
 &  &  &  & 3 & 1 & -4.967570 & 4.354450\\
 &  &  &  & 3 & 2 & -4.788530 & 4.056980\\
 &  &  &  & 3 & 3 & -3.484330 & 3.682040\\
\end{tabular}
\end{ruledtabular}
\end{table}

\newpage

\begin{table}
\begin{ruledtabular}
\begin{tabular}{cccccccc}
 &  &  & continued &  &  &  & \\ \hline
$m_{1}$ & $m_{0}$ & $V_{0}$ & $S_{0}$ & $n$ & $\ell$ &
$E_{a}$ & $E_{p}$\\ \hline
0.1 & 5 & -1 & 2 & 1 & 0 & -3.973190 & 4.994930\\
 &  &  &  & 1 & 1 & -4.456350 & 4.991980\\
 &  &  &  & 2 & 0 & -4.733050 & 4.950930\\
 &  &  &  & 2 & 1 & -4.901740 & 4.879150\\
 &  &  &  & 2 & 2 & -4.997830 & 4.727350\\
 &  &  &  & 3 & 0 & -4.980140 & 4.789140\\
 &  &  &  & 3 & 1 & -4.999330 & 4.673420\\
 &  &  &  & 3 & 2 & -4.934260 & 4.460740\\
 &  &  &  & 3 & 3 & -4.749570 & 4.159960\\ \hline
1 & 5 & -5 & 10 & 1 & 0 & -1.8565680 & 4.9226060\\
 &  &  &  & 1 & 1 & -2.060403 & 4.948111\\
 &  &  &  & 2 & 0 & -3.156077 & 4.996077\\
 &  &  &  & 2 & 1 & -3.292089 & 4.989522\\
 &  &  &  & 2 & 2 & -3.537530 & 4.968551\\
 &  &  &  & 3 & 0 & -4.025257 & 4.881421\\
 &  &  &  & 3 & 1 & -4.115249 & 4.856386\\
 &  &  &  & 3 & 2 & -4.276053 & 4.801841\\
 &  &  &  & 3 & 3 & -4.475750 & 4.710567\\
\end{tabular}
\end{ruledtabular}
\end{table}

\end{document}